\documentstyle[aps,prl,epsf,twocolumn,floats]{revtex}
\begin{document}
\draft
\twocolumn[\hsize\textwidth\columnwidth\hsize\csname 
@twocolumnfalse\endcsname
\title{\bf Spin Injection into a Luttinger Liquid}
\author{Qimiao Si}
\address{Department of Physics, Rice University, Houston, 
TX 77251-1892}

\maketitle
\begin{abstract}

We study the effect of spin injection into a Luttinger liquid.
The spin-injection-detection setup of Johnson and Silsbee
is considered; here spins injected into the Luttinger
liquid induce, across an interface with a ferromagnetic metal,
either a spin-dependent current ($I_s$) or
a spin-dependent boundary voltage ($V_s$).
We find that the spin-charge separation nature of the
Luttinger liquid affects $I_s$ and $V_s$ in a very 
different fashion. In particular, in the Ohmic regime,
$V_s$ depends on the spin transport properties of 
the Luttinger liquid in essentially the same way 
as it would in the case of a Fermi liquid.
The implications of our results for
the spin-injection-detection experiments
in the high $T_c$ cuprates are discussed.

\end{abstract}
\pacs{PACS numbers: 71.10.Hf, 73.40. -c, 71.27. +a, 72.15.Gd}
]

\narrowtext

Spin-charge separation has long been proposed to describe 
the normal state of the high $T_c$ cuprate 
superconductors\cite{Anderson1}. Existing experimental results
cited as evidences for spin-charge separation are mostly on
transport properties\cite{Anderson2}.
Since only charge transport properties have so far been measured,
the inference about the coupling, or 
lack thereof, between the underlying spin and charge
excitations is indirect.
It would appear natural that spin transport, when combined 
with charge transport, should be useful in this
context. Indeed we have recently proposed to probe spin-charge 
separation using a comparison between the temperature 
dependence of the yet-to-be-measured spin resistivity and 
that of the known electrical resistivity\cite{Si}.
Several factors point to the feasibility of
experimentally measuring spin transport
in the cuprates using the spin-injection-detection
technique\cite{Johnson1,Johnson2,Johnson3}.
First of all, progresses in the preparation of 
the manganite-cuprate heterostructures appear to have led
to the first demonstration of spin injection into
the cuprates (albeit in the superconducting
state)\cite{Goldman,Venkatesan}. Secondly,
the spin-diffusion length in the cuprates has been 
estimated to fall in the range required by this
technique\cite{Si}. 

In light of the new experiments on the high $T_c$ cuprates,
it is important to understand how spin injection into a non-Fermi
liquid differs from spin injection into a Fermi liquid. 
The effects of spins injected into a non-interacting electron system
has been studied in the past by Johnson and Silsbee
and others\cite{Johnson1,Johnson2,Johnson3,Fert,Hershfield},
following the initial proposals for spin injection and
detection\cite{Aronov,Silsbee}.

Here we address the influence of spin-charge separation
in the bulk metal on the boundary voltage/current measured 
in a spin-injection experiment. For definiteness, 
the one-dimensional Luttinger liquid\cite{Voit} 
is used as a prototype for a spin-charge-separated metal.
Crucial for our analysis is the fact that, the interface 
transport involves the binding of spinons and holons which 
then tunnel as a whole from the Luttinger liquid into 
the ferromagnetic metal. We find that the spin-charge-separation
nature of the Luttinger liquid affects the boundary voltage 
and boundary current in a very different fashion. 

\begin{figure}[t]
\centerline{
\vbox{\epsfxsize=80mm \epsfbox{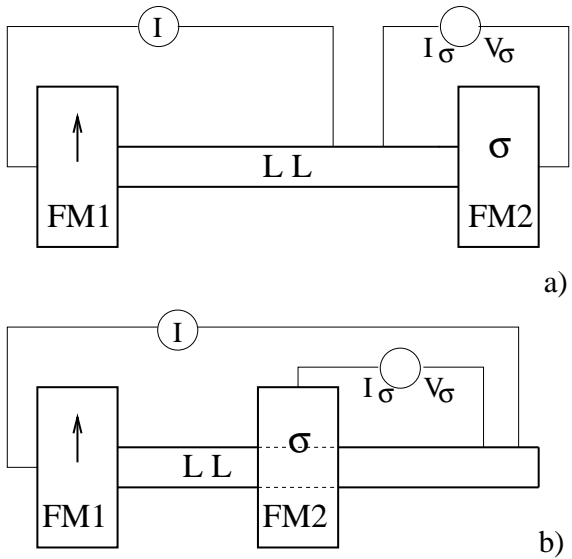}}}
\caption{
Two possible set ups for spin-injection-detection experiment
on a one-channel
Luttinger liquid (LL). The magnetization of the ferromagnetic metal
FM2 is either parallel ($\sigma=\uparrow$) or antiparallel
($\sigma=\downarrow$) to that of the ferromagnetic metal FM1.
$I$ is the injection current. $I_{\sigma}$ ($V_{\sigma}$) is the current
(boundary voltage) induced across the LL-FM2 interface when the circuit
is closed (open). The two interfaces in a) are separated by 
a distance $d$.
}
\label{fig:setup}
\end{figure}

Figs. \ref{fig:setup}a) and \ref{fig:setup}b) illustrate
two specific geometries involving a 
single channel Luttinger liquid (LL).
The Luttinger liquid is in contacts with 
two itinerant ferromagnets, FM1 and FM2. 
The magnetization 
of FM1 is chosen as the $\uparrow$ direction.
The magnetization 
of FM2 is either parallel ($\sigma=\uparrow$) or antiparallel
($\sigma=\downarrow$) to that of FM1.
One passes an electrical current ($I$) across the FM1-LL
interface. This current serves to inject non-equilibrium
magnetization into the Luttinger liquid. 
For a given $\sigma$, $I_{\sigma}$ represents the induced current 
across the LL-FM2 interface in a closed circuit and $V_{\sigma}$ is 
the induced boundary voltage ($V_{\sigma}$) in an open circuit. 
The spin-dependent current, $I_s$,
is defined  as the difference between the induced current
when the magnetizations of the two ferromagnets are 
in parallel and that when they are antiparallel.
Likewise, the spin-dependent voltage, $V_s$,
is the difference between the induced boundary 
voltages in the corresponding two cases.
The setups illustrated in Figs. \ref{fig:setup}a) and
\ref{fig:setup}b) differ in two regards.
First of all, while the Luttinger liquid is in point contacts with 
both of the ferromagnets in Fig. \ref{fig:setup}a), 
in Fig. \ref{fig:setup}b) it is in contact with FM2 over 
an extended spatial region. Secondly, unlike in 
Fig. \ref{fig:setup}a) the injection and detection
loops in Fig. \ref{fig:setup}b) are closed through contacts 
with LL far away both FM1 and FM2.
The setup of Fig. 1b) is perhaps easier to implement experimentally.
On the other hand, that of Fig. 1a) is easier to analyze theoretically.
To illustrate the basic principle, in the rest of this paper
we will focus on Fig. \ref{fig:setup}a).

The Hamiltonian of the Luttinger liquid can be written as
\begin{eqnarray}
H_{lut} = && H_{\rho} + H_{s} + H'\nonumber\\
H_{\rho} = && { 1 \over 2\pi} v_{\rho} \int dx ~~[ K_{\rho} (\pi 
\Pi_{\rho})^2 + {1 \over K_{\rho}} (\partial_x \phi _{\rho})^2] \nonumber\\
H_{s} = && { 1 \over 2\pi} v_{s} \int dx ~~[ K_{s} (\pi \Pi_{s})^2
+ {1 \over K_{s}} (\partial_x \phi _{s})^2]
\label{hamlut0}
\end{eqnarray}
where $H_{\rho}$ and $H_{s}$ are respectively the Hamiltonian for
the charge ($\rho$) and spin ($s$) bosons,
$\phi_{\rho}$ and $\phi_{s}$; $\Pi_{\rho}$ and $\Pi_{s}$ 
are the corresponding conjugate momenta. The charge and spin 
velocities, $v_{\rho}$ and $v_{s}$, and Luttinger liquid parameters,
$K_{\rho}$ and $K_{s}$, are determined by the forward 
scattering interactions. We consider the case when the
spin-rotational invariance is preserved so that $K_s=1$.

We focus on the regime where spin transport inside the Luttinger liquid
is diffusive, with a spin diffusion constant $D_s$. The diffusive transport
is the result of the dissipative terms in the Hamiltonian, $H'$,
which also lead to a finite spin relaxation time $T_1$.
The precise form of $H'$ is however unimportant for our purpose
in this paper and is left unspecified here. 
(It will, of course, determine the specific temperature dependences
of $D_s$ and $T_1$.) To simplify the discussion,
we assume that the FMs are half-metallic ferromagnets. We will also
neglect the electron interactions inside the 
ferromagnets\cite{magnon}. The free electron Hamiltonians 
for FM1 and FM2 are respectively
\begin{eqnarray}
H_1 = \sum_k \epsilon_k^1 c_{k\uparrow}^{\dagger} c_{k\uparrow}
\label{ham.left}
\end{eqnarray}
and 
\begin{eqnarray}
H_2 = \sum_k \epsilon_k^2 c_{k,\sigma}^{\dagger} c_{k,\sigma}
\label{ham.right}
\end{eqnarray}
where $\epsilon_k^1$ and $\epsilon_k^2$ are the corresponding
energy dispersions.

Since we will consider only the cases when the magnetizations of
the two ferromagnets are either parallel or anti-parallel with
each other, we need only the kinetic equation
for the longitudinal component of the magnetization. We introduce
$m(x)$ to denote the deviation of the steady state
magnetization density from the corresponding equilibrium
value. The FM1-LL and LL-FM2 interfaces are located at $x=0$ and $x=d$,
respectively. The non-equilibrium magnetization density $m(x)$ 
satisfies the following\cite{transverse}
\begin{eqnarray}
-D_s {\partial ^ 2 m \over \partial x^2} = -{ {m(x)} \over T_1}
\label{kinetic}
\end{eqnarray}
where $T_1$ is the longitudinal spin relaxation time.
The boundary conditions will depend on the details of the interface.
Leaving more general cases for elsewhere\cite{Si2},
we assume that no spin-flip scattering exists at the interface.
In this case, the spin current $j_s$ is conserved across 
the interface.
Given that $j_s (x=0^-) = \mu_B I/e $, where $\mu_B$ is the 
Bohr magneton and $e$ the electron charge, the boundary 
condition at the FM1-LL interface is,
\begin{eqnarray}
-D_s \partial m / \partial x |_{x=0} = \mu_B I/e
\label{bc.spincurrent}
\end{eqnarray}
Likewise, at the LL-FM2 interface,
\begin{eqnarray}
-D_s \partial m / \partial x |_{x=d} = \mu_B I_{\sigma}/e
\label{bc.spincurrent2}
\end{eqnarray}

The induced current $I_{\sigma}$ depends on the 
drop of the non-equilibrium magnetization across the LL-FM2
interface, which is equal to $\Delta m = m(d)$.
We separately discuss the closed circuit and open circuit cases.

Consider, first, the case of a closed circuit. 
The drop of non-equilibrium magnetization,
$\Delta m$, leads to a drop in the
effective magnetic field,
\begin{eqnarray}
\Delta H = \Delta m / \chi
\label{Delta.H}
\end{eqnarray}
across the LL-FM2 interface. Here, 
 $\chi$ is the uniform spin susceptibility of the 
Luttinger liquid. $\Delta H$ provides a driving force
for spinons to move across the interface. 
Since only electrons can tunnel across the barrier,
spinons bind with holons and move across the interface 
as a whole. A finite electrical current, $I_{\sigma}^{closed}$, 
is then induced by $\Delta H$.
To calculate $I_{\sigma}^{closed}$, 
we follow the general
procedure of Kane and Fisher\cite{Kane-Fisher} and 
integrate out all the degrees of freedom of
the Luttinger liquid except at the site of contact,
$x=d$. This leads to an effective action entirely
determined by the boson field $\phi_{\rho} \equiv
\phi_{\rho}(d)$ and $\phi_{s} \equiv \phi_{s}(d)$:
\begin{eqnarray}
S_{site} = K_{\rho} {1 \over \beta} \sum_{\omega_n} |\omega_n| 
|\phi_{\rho} (\omega_n) | ^ 2
+ 
{1 \over \beta} \sum_{\omega_n} |\omega_n| 
|\phi_{s} (\omega_n) | ^ 2
\label{action.site}
\end{eqnarray}
where $\beta$ is the inverse of temperature and $\omega_n$
the bosonic Matsubara frequencies.
In deriving this on-site action, we have neglected the effect of the
non-equilibrium magnetization, $m(x)$. This is appropriate
for the Ohmic regime $k_BT > \mu_B m/\chi $. 
We are now faced with a problem of one\cite{multi-mode} retarded
impurity -- whose dynamics is controlled by 
Eq. (\ref{action.site}) -- coupled to
a three dimensional ferromagnetic metal.
This effective impurity problem is illustrated in Fig. 2.
The impurity problem can also be written in a Hamiltonian form,
by introducing a fictitious bosonic bath.

\begin{figure}[t]
\centerline{
\vbox{\epsfxsize=80mm \epsfbox{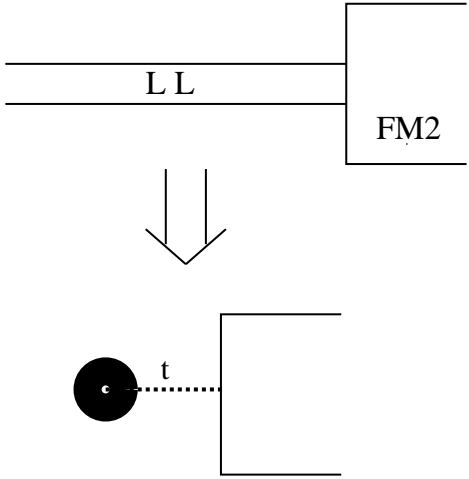}}}
\caption{
Reduction of the coupled LL-FM2 problem to a retarded impurity,
described by the action given in Eq. (\ref{action.site}),
coupled to a three-dimensional ferromagnetic metal.
The coupling is through the tunneling matrix $t$.
}
\label{fig:impurity}
\end{figure}

From the one-particle tunneling Hamiltonian, one
can construct 
respectively the charge current operator $J$ and spin current
operator $J_M$. They are as follows,
\begin{eqnarray}
J/e = &&J_M / {\mu_B} \nonumber \\
 = &&{t \over \sqrt{2\pi a}}
[F_{\sigma -}^{\dagger} 
{\rm e}^{-i{ {\phi_{\rho} + \theta_{\rho}} \over \sqrt{2}} 
-i\sigma {{\phi_{s} + \theta_{s}} \over \sqrt{2}}} \nonumber\\
&&+
F_{\sigma +}^{\dagger} 
{\rm e}^{-i{{-\phi_{\rho} + \theta_{\rho}} \over \sqrt{2}}
-i\sigma {{-\phi_{s} + \theta_{s}} \over \sqrt{2}}}]
c_{\sigma} -H.c.
\label{current}
\end{eqnarray}
where $F_{\sigma \pm}^{\dagger}$ are the Klein operators for
the left and right moving branches and $a$ is a lattice
cutoff\cite{Voit,Klein}. Here, $c_{\sigma}$ is the 
 annihilation operator for the Wannier orbital of the 
conduction electrons at the contact, and $t$ is the tunneling
matrix. That the interface charge and spin current operators
are simply related to each other\cite{half-metallic} 
in spite of the separated
spin and charge excitations in the bulk Luttinger liquid
reflects the simple physics that, only a bare electron
tunnels across the interface. 

We can now calculate $I_{\sigma}^{closed}$ in the Ohmic regime 
using the Kubo formalism,
\begin{eqnarray}
I_{\sigma}^{closed} = (-\Delta H) 
\lim_{\omega \rightarrow 0} {{-{\rm Im}{\pi_{JJ_M} (\omega+i0^+) }} 
\over \omega }
\label{Iind.1}
\end{eqnarray}
where $\pi_{JJ_M}$ is the charge current-spin 
current correlation function.
Similar to the Kane-Fisher problem,
for repulsive interactions $K_{\rho} <1$,
the tunneling term is an irrelevant coupling 
in the renormalization group sense.
We can then calculate $\pi_{JJ_M}$ 
perturbatively in $t$. The result is as follows,
\begin{eqnarray}
I_{\sigma}^{closed} = C (k_BT)^{1/2K_{\rho} - 1/2}
e \mu_B \Delta m /\chi
\label{Iind1}
\end{eqnarray}
where $C=c N_F t^2 / W^{1/2K_{\rho}+1/2}$. Here $N_F$ is the density
of states of FM2 at the Fermi energy, $W$ is a typical bare energy
scale associated with the electrons in the Luttinger liquid, and $c$ 
is a constant of order unity. The induced spin-dependent current 
across the LL-FM2 interface,
$I_s = I_{\uparrow}^{closed} - I_{\downarrow}^{closed}$ 
can now be determined, from Eqs.
(\ref{kinetic},\ref{bc.spincurrent},\ref{bc.spincurrent2},\ref{Iind1}).
The result is as follows,
\begin{eqnarray}
I_{s} / I = C (k_B T)^{1/2K_{\rho} - 1/2}
{ \mu_B^2 \over  \chi} {T_1 \over \delta_s \sinh (d / \delta_s)} 
\label{Iind.cc}
\end{eqnarray}
where $\delta_s = \sqrt{D_s T_1}$ is the spin-diffusion
length of the Luttinger liquid. 

We now turn to the open circuit case. Here, in order to balance the 
current induced by the magnetization drop, a boundary voltage,
$V_{\sigma}$, develops across the LL-FM2 interface. 
The induced current in this case is
\begin{eqnarray}
I_{\sigma}^{open} = &&
(-\Delta H )
\lim_{\omega \rightarrow 0} {{-{\rm Im}{\pi_{JJ_M} (\omega+i0^+)}
 \over \omega}} \nonumber\\
&&+ V_{\sigma} \lim_{\omega \rightarrow 0}
{{-{\rm Im}{\pi_{JJ} (\omega+i0^+)}}
\over \omega}
\label{Iind.2}
\end{eqnarray}
which leads to
\begin{eqnarray}
I_{\sigma}^{open} = C (k_BT)^{1/2K_{\rho} - 1/2} e
(\mu_B \Delta m/\chi + e V_{\sigma})
\label{Iind2}
\end{eqnarray}
Setting $I_{\sigma}^{open}=0$ in Eq. (\ref{Iind.2}),
and combining with
Eqs. (\ref{kinetic},\ref{bc.spincurrent},\ref{bc.spincurrent2}), we
arrive at the following expression for the spin-dependent boundary
voltage, $V_s = V_{\uparrow}- V_{\downarrow}$:
\begin{eqnarray}
V_s / I = {\mu_B^2  \over e^2  \chi } {T_1 \over {\delta_s \sinh (d /
\delta_s )}}
\label{Vs}
\end{eqnarray}

Eqs. (\ref{Iind.cc},\ref{Vs}) are the main results of this work.
Several comments are in order.

First of all, it is instructive to see how our results reduce
to those for free 
electrons\cite{Johnson1,Johnson2,Johnson3,Fert,Hershfield}
when electron interactions are reduced to zero.
For non-interacting electrons, 
spin diffusion $D_s$ is reduced to the usual electron
diffusion constant $D = v_F^2 \tau$, where $v_F$ is the Fermi
velocity and $\tau$ the transport scattering time.
In addition, $\chi = \mu_B^2 N_F^P$, where $N_F^P$ is 
the density of states at the Fermi energy.
Straightforward manipulation leads to 
$V_s/I = \rho \delta_s^0 / \sinh (d/\delta_s^0)$,
where $\rho$ is the electrical resistivity
of the bulk metal and $\delta_s^0=\sqrt{D T_1}$.
In addition, for non-interacting electrons, $K_{\rho}=1$,
and our expression for $I_s$ is reduced to 
$I_s/I = (e^2N_F N_F^P t^2) \rho \delta_s / \sinh (d/\delta_s)$.

Secondly, we note that for the Luttinger liquid,
the temperature dependence of $I_s$ 
is not solely determined by that of the bulk 
spin diffusion and relaxation properties of 
the Luttinger liquid.
There is an additional temperature-dependent factor, with
a power which explicitly depends on the Luttinger liquid 
parameter $K_{\rho}$. This additional temperature-dependent
factor, however, cancels out in $V_s$. This last result
can ultimately be traced to the fact
that the interface charge current and spin current are
directly related to each other,
in spite of the spin-charge separation nature of the 
bulk Luttinger liquid. On this ground, we expect the expression
for $V_s$ to be valid very generally,
so long as one particle processes dominate the interface 
transport and no strong-coupling (in $t$) 
phenomena\cite{Fradkin} take place.
The boundary voltage $V_s$ is hence more useful than 
$I_s$ for the purpose of extracting bulk spin transport
properties of strongly correlated metals, including 
the high $T_c$ cuprates.

The general expression for the spin-dependent boundary voltage
suggests the following procedure to measure the temperature
dependence of the spin transport scattering rate, 
$1/\tau_{tr,spin}$, of correlated metals. The latter is
the quantity of interest in probing spin-charge
separation\cite{Si}.
In strongly interacting metals, it is likely that
the dissipations for both the spin current and total spin
come primarily from electron-electron interactions.
The spin relaxation time and spin transport relaxation
rate are then proportional to each other:
${1 \over T_1} 
\approx (\lambda_{so})^2 {1 \over \tau_{tr,spin}}$, where 
$\lambda_{so}$ is the dimensionless spin-orbit coupling 
constant. When the sample thickness $d$ is small compared to
the spin diffusion length $\delta_s$,
the temperature dependence of $\chi V_s/I$ is 
then directly proportional to the temperature dependence
of $\tau_{tr,spin}$. On the other hand, for thickness much larger
than the spin diffusion length, the temperature dependence
of $\ln (\chi V_s / I )$ is directly proportional to that
of $1/\tau_{tr,spin}$.
This procedure does not require measurement in a series
of samples of different thicknesses -- as was necessary in
the case of simple metals\cite{Johnson2} -- and may be
experimentally easier to implement.

To summarize, we have studied the effects of spin injection
into a Luttinger liquid. Our conclusion that the 
temperature dependence of the boundary voltage depends on 
the bulk spin-transport properties only is expected
to be generally applicable and of direct relevance 
to the spin-injection-detection experiments in the cuprates. 

I would like to thank N. Andrei, M. Johnson, Q. Li and
A. J. Rimberg for useful discussions, and the Aspen Center for Physics
and the CM/T group at NHMFL/FSU for hospitality and support
during different stages of this work. The work has also been
supported by NSF Grant No. DMR-9712626,
a Robert A. Welch Foundation grant and an A. P. Sloan Fellowship.


\begin{references}
\bibitem{magnon}Having super-ohmic spectral functions, 
the spin waves in the ferromagnets
are unimportant for our analysis.
\bibitem{half-metallic} When FM2 is not half-metallic,
the relationship between $J$ and $J_M$ will be more complex.
The conclusions are, however, essentially the same 
as give here\cite{Si2}.
\end{references}
\end{document}